\documentclass[preprint,preprintnumbers,amsmath,amssymb]{revtex4}
\usepackage{graphicx}
\usepackage{dcolumn}
\usepackage{bm}
\begin{document}
\title{ Fermat's principle in quantum gravitational optics}
\author{ N.~Ahmadi$^{a}$\footnote{
Electronic address:~nahmadi@ut.ac.ir}, S.~Khoeini-Moghaddam$^{b ,
c}$\footnote{Electronic address:~saloumeh@mehr.sharif.edu} and
M.~Nouri-Zonoz $^{a , c}$ \footnote{ Electronic
address:~nouri@theory.ipm.ac.ir, Corresponding author}}
\address{$^{a}$ Department of Physics, University of Tehran, North Karegar Ave., Tehran 14395-547, Iran. \\
$^{b}$ Department of Physics, Sharif
University of Technology, P O Box 11365-9161 Tehran, Iran. \\
$^{c}$ Institute for studies in theoretical physics and mathematics, P O Box 19395-5531 Tehran, Iran.}
\begin{abstract}
Interactions incorporating the vacuum polarization effects in
curved backgrounds modify the null cone structure in such a way
that the photon trajectories would not be the space-time geodesics
anymore. The gravitational birefringence introduced as a direct
consequence of these effects, will allow shifts in the photon
velocities leading to polarization dependent superluminal
propagation. Taking these effects into account we study Fermat's
principle in the context of the 1+3 (threading) formulation of the
space-time decomposition. We find an expression for the modified spacetime 
refractive index and show it is proportional to the 
light cone correction to the first order. Consequences of this modification on
polarization sum rules and spatial light paths are considered.
\end{abstract}
\maketitle
\section{Introduction}
Quantum Gravitational Optics (QGO) is the term coined for the
studies incorporating photon propagation in curved space-times in
the presence of QED vacuum polarization in the lowest (leading)
order \cite{1,2}. The quantum corrections introduce null cone
modifications which in general alter the propagation
characteristics of photons leading to cases in which they could
travel at speeds greater than unity. It is expected that the more
complex the space-time is, the more complex should be its
curvature coupling to these quantum effects. In line with the
above expectation, this phenomenon has been studied in
Schwarzschild,  Robertson-Walker and gravitational wave
backgrounds \cite{1} as well as in Reissner-Nordstrom and Kerr
black hole space-times \cite{3,4}. It is found that except for the
case of de Sitter space there may be certain directions and
polarizations in which photons travel at speeds greater than
unity. It is also shown that in the case of static topological
black holes, the velocity shift of photons is the same as in the
Reissner-Nordstrom case and the vacuum polarized photons are not
sensitive to the asymptotic behavior or topological structure of
space-time while for radiating topological black hole backgrounds,
the light cone condition and velocity shifts depend on the
topological structure \cite{5}. In a recent paper, taking the
vacuum polarization into account the propagation of a bundle of
rays in a background gravitational field has been studied through
the perturbative deformation of the Raychaudhuri equation
\cite{6}. However, there are other important phenomena predicted
in the framework of general relativity (GR) while semiclassical
corrections such as the QED interactions  mentioned above are
taken into account. Despite the fact that  QGO corrections are
expected to be small, the comparison between the classical (GR)
and semiclassical (QGO) theories can be conducted on the basis of
higher order effects which deserve attention from both the
theoretical and observational points of view.
In the realm of photon propagation, the observational testbeds are usually provided in the context of
gravitational lensing through the impressive development of technical capabilities. Hence a special treatment of lensing phenomena could be provided through the consideration of these
quantum effects. Following a particular version of Fermat's principle for stationary space-times, we obtain an equation
for the spatial light paths which demonstrates why, for a QED photon, a curved background acts as if it is an anisotropic medium with its
own polarization dependent refractive index.

In what follows, first we give a brief account of QGO. Then, based on Fermat's principle in the context of $1+3$ (threading) formulation of space-time decomposition,
it is shown how the basic modifications introduced by the theory 
lead to the corrections on the photon spatial trajectories and the spacetime
refractive index. After introducing an intriguing expression for an {\it effective local refractive index} in section $3$, it is shown in section $4$ that its first order correction is proportional to the null cone modification thereby leading to two new theorems on a)-polarization summed shift of refractive indices and b)-photons travelling normal to the horizon.
In section $5$ using the results of Tsai and Erber \cite{Tsai} on the refractive indices of photons of arbitrary frequency propagating in a background magnetic field, gravitational and electromagnetic birefringence are compared. In section $6$ the equation of spatial light paths in the presence of the above QGO effects is given. The results are summarized and discussed in the last section.
\section{Vacuum polarized photon propagation in a curved background}
Vacuum polarization is an effect in which the photon exists as a
virtual $e^{+}e^{-}$ pair for a short time. This virtual
transition assigns photons with an effective size of
${\mathcal{O}}(\lambda_c)$, where $\lambda_c$ is the Compton
wavelength of the electron \cite{1}. The photon propagation will
be affected by gravity if the scale of space-time curvature $L$ is
comparable to $\lambda_c$. Ignoring the vacuum polarization, the
equivalence principle leads to the photon propagation at the speed
of light on the space-time null geodesics. But when the pair
production is taken into account, the equivalence principle is
violated in such a way that the photon trajectories are modified
and superluminal (photon) propagation becomes a possibility. All
this could happen without necessarily breaking down the causality
\cite{2}. Considering the vacuum polarization, the first order
correction to the electromagnetic action
\begin{equation}
W_0=-\frac{1}{4}\int d^4x\sqrt{-g}F_{\mu\nu}F^{\mu\nu}
\end{equation}
is given by
\begin{eqnarray}\label{Drumhath}
W_1&=&\frac{1}{m_{e}^{2}}\int
d^4x(-g)^{\frac{1}{2}}(aRF_{\mu\nu}F^{\mu\nu}+bR_{\mu\nu}F^{\mu\sigma}F^{\nu}_{\sigma}+cR_{\mu\nu\sigma\tau}F^{\mu\nu}F^{\sigma\tau}
+dD_{\mu}F^{\mu\sigma}D_{\nu}F^{\nu}_{\sigma})
\end{eqnarray}
in which
\begin{eqnarray}\label{coefficients}
a=-\frac{1}{144}\frac{\alpha}{\pi}\hspace{1.5cm}b=\frac{13}{360}\frac{\alpha}{\pi}\hspace{1.5cm}c=-\frac{1}{360}\frac{\alpha}{\pi}
\hspace{1.5cm}d=-\frac{1}{30}\frac{\alpha}{\pi}
\end{eqnarray}
are perturbative coefficients of ${\mathcal{O}}(\alpha)$, $m_{e}$
is the electron mass and $\alpha$ is the fine structure constant.
Expression (\ref{Drumhath}) is called {\em Drummond-Hathrell}
action \cite{1}. The first three terms show the influence of
curvature while the last one,     representing the off-mass-shell
effects (in the vacuum polarization), exists even in the flat
spacetime \footnote {The coefficients $a$, $b$ and $c$ may be
obtained from the coupling of a graviton to two on-mass-shell
photons in the flat-space limit. The term with coefficient {\it d}
is neglected in the following study because it is already of
${\mathcal{O}}(\alpha^2)$.}. The effective equation of motion can
be obtained by varying $W=W_0+W_1$ with respect to $A_\nu$ as
follows,
\begin{equation}\label{equation of motion}
D_{\mu}F^{\mu\nu}-\frac{1}{m_{e}^{2}}[2bR_{\mu\sigma}D^{\mu}F^{\sigma\nu}+4cR_{\mu\hspace{3mm}\sigma\tau}^{\hspace{2mm}\nu}D^{\mu}F^{\sigma\tau}]=0
\end{equation}
in which only terms of first order in $\alpha$ and
$\left(\lambda_c/L\right)^{2}$ are kept.
Now we use the geometrical optics approximation,
in which the electromagnetic field is written as a slowly-varying
amplitude and a rapidly-varying phase through a small
parameter $\epsilon$, i.e, $ (A_\mu+i\epsilon
B_\mu+\cdots)\exp(i\frac{\vartheta}{\epsilon})$, where $\vartheta$
is a real scalar field. In geometric optics, the wave vector is defined as the gradient of the phase, i.e, $k_\mu=
\partial_\mu\vartheta$ and the polarization vector is introduced through $ A_\mu=Aa_\mu$, where $A$ is the amplitude and
$a_\mu$ is the polarization vector.
Rewriting the equation of motion (\ref{equation of motion}) as an equation for the polarization vector we have \cite{8},
\begin{eqnarray}\label{pol}
k^2a_\sigma+\frac{2b}{m_{e}^{2}}\left[R_{\mu\nu}(k^\mu
k^\nu a_\sigma-k^\mu k_\sigma a^\nu)\right]+\frac{8c}{m_{e}^{2}}R_{\mu\sigma\nu\tau}k^\mu k^\nu
a^\tau=0,
\end{eqnarray}
From which the light cone condition at ${\mathcal{O}}(\epsilon^{-1})$ is given by
\begin{eqnarray}\label{light cone}
k^2+\frac{2b}{m_{e}^{2}}R_{\mu\nu}k^\mu
k^\nu-\frac{8c}{m_{e}^{2}}R_{\mu\sigma\nu\tau}k^\mu k^\nu a^\tau
a^\sigma=0,
\end{eqnarray}
in which $a^\mu a_\mu =-1$. Since this effective equation of motion is homogeneous and quadratic in $k^\mu$ it can be
written as
\begin{equation}\label{eff}
{\mathcal G^{\mu\nu}}k_\mu k_\nu=0,
\end{equation}
in which $\mathcal G^{\mu\nu}$ could be defined as an effective (physical) metric whose null geodesics represent the
modified photon trajectories. In other words at this order of approximation there is no dispersion
 and consequently the phase and group velocities are equal and given by
\begin{equation}
v_{ph}=\frac{k^0}{|k|}
\end{equation}
From (\ref{eff}) it could be seen that there is a possibility for the modified null cone to lie outside the original
light cone leading to superluminal photon velocities as well as modified trajectories. It is this feature of QGO that will
be the main focus of our study through Fermat's principle in the forthcoming  sections.

\section{Fermat's principle in QGO}
One way to look at the nature of the modified null trajectories in QGO is through the fundamental definition of null rays
via Fermat's principle. Within the assumptions of geometrical optics, the trajectory of a light ray $\gamma$ from a
source $P$ to an observer on his world line $\cal L$, in a space-time $({\cal M}, g_{\mu\nu})$, can be characterized by
the following relativistic version of Fermat's principle \cite{9}:
{\it Signal's arrival time is stationary with
respect to the first order variations of $\gamma$ within the set of smooth null curves from $P$ to $\cal L$ }.\\
In stationary space-times Fermat's principle takes a version of
particular interest. To invoke this version of the principle, we
employ the $1+3$ (threading) formulation of space-time
decomposition in which the space-time metric can be written in the
following form \cite{10}
\begin{equation}
ds^{2}=h\left(dx^0-A_{i}dx^{i}\right)^2-dl^2
\end{equation}
where
\begin{equation}
A_{i}\equiv g_i =-\frac{g_{0i}}{g_{00}}\ \ , \ \ h=g_{00}
\nonumber
\end{equation}
and
\begin{equation}
dl^2=\gamma_{ij}dx^{i}dx^{j}=\left(-g_{ij}+\frac{g_{0i}g_{0j}}{g_{00}}\right)
dx^{i}dx^{j}\ \ , \ \ i,j=1,2,3 \nonumber
\end{equation}
is the spatial distance in terms of the 3-dimensional metric
$\gamma_{ij}$ of the three dimensional space. Using the identity
$g^{\mu\nu}g_{\nu\sigma}=\delta^{\mu}_{\sigma}$ it could be shown
that the following relations also hold
\begin{equation}
\gamma^{ij} = -g^{ij} \ \ , \ \ g^i = -g^{0i} \ \ , \ \ g^{00} =
\frac{1}{g_{00}}-g_i g^i \label{relation}
\end{equation}
Using the classical version of Fermat's principle (i.e at $O\left(\alpha^{0}\right)$) \cite{10}, we obtain its QGO version
for the propagation of null
rays in an stationary gravitational field, taking into account the above vacuum polarization effects. Fermat's principle
 states
\begin{equation}
\delta\int{k_{i}dx^{i}=0},
\end{equation}
where the integral is taken along the spatial projection of the physical rays which are to be varied with fixed endpoints. Rewriting the light cone
equation (\ref{light cone}) in the form,
\begin{eqnarray}
&&{\tilde k}^2-A_{\mu\nu} {\tilde k}^{\mu} {\tilde k}^{\nu}=0\nonumber\\
&& A_{\mu\nu}\equiv
\frac{1}{m_{e}^{2}}\left(8cR_{\mu\sigma\nu\tau}a^{\tau}a^{\sigma}-2b
R_{\mu\nu}\right)
 ,\label{null cone0}\end{eqnarray}
 we obtain
 \begin{equation}
h\left({\tilde k}^{0}-g_{i}{\tilde
k}^{i}\right)^2-\gamma_{ij}{\tilde k}^{i}{\tilde k}^{j}=A_{00}
{\tilde k}^{0} {\tilde k}^{0}+2A_{0i} {\tilde k}^{0} {\tilde
k}^{i}+A_{ij} {\tilde k}^{i} {\tilde k}^{j} .\label{null cone}
\end{equation}
In which ${\tilde k}$ stands for the modified wave vector to the
first order in $\alpha$. Noting that the vector $k^{i}$ must have
the direction of the vector $dx^{i}$ for the classical photon
(zeroth-order of the above equation), i.e, $k^i =
\frac{k_0}{\sqrt{h}} \frac{dx^i}{dl}$ we find its correction of
$O\left(\alpha\right)$ by substituting the classical values for
$k^i$ and $k^0$ in the right hand side of (\ref{null cone}), and
the conjectured modified wave vector ${\tilde k}^i =
\frac{k_0}{\sqrt{h}} \frac{dx^i}{dl}+B^i$ in its left hand side so
that it reduces to
\begin{equation}
\frac{{k_0}^2}{h}-\gamma_{ij}({\frac{k_0}{\sqrt{h}}
\frac{dx^i}{dl}+B^i})({\frac{k_0}{\sqrt{h}}
\frac{dx^j}{dl}+B^j})=A_{00}(k^0)^2+2A_{0i}{
k}^{0}\frac{k_0}{\sqrt{h}} \frac{dx^i}{dl}+A_{ij}
\frac{{k_0}^2}{h} \frac{dx^i}{dl} \frac{dx^j}{dl} ,\label{null
cone1}
\end{equation}
where we have used the fact that
\begin{equation}
k_0=g_{0i}k^i=h(k^0-g_i k^i )=h({\tilde k}^0-g_i
{\tilde k}^i) \label{wave}
\end{equation}
is the constant frequency along the ray. After a few lines of simple manipulations it is found that
\begin{equation}
B^i = -\frac{\sqrt{h}}{2k_0}\left(A_{00}(k^0)^2 \frac{dx^i}{dl} +
2A_{0j} \gamma^{ji} \frac{k_0{k^0}}{\sqrt{h}} + \frac{{k_0}^2}{h}
A_{kj}\gamma^{ji}\frac{dx^k}{dl}\right)
\end{equation}
and consequently substituting for $k^0$ from (\ref{wave}) we end up with
\begin{equation}
{\tilde k}^{i}=\frac{k_0}{\sqrt{h}}
\frac{dx^i}{dl}\left(1-\frac{A_{00}}{2}D^2\right)-\frac{k_0}{2\sqrt{h}}\left(
2A_{0j} \gamma^{ji}D + A_{kj}\gamma^{ji}\frac{dx^k}{dl}\right)
\end{equation}
in which $D=\frac{1}{\sqrt{h}} + g_i \frac{dx^i}{dl}$. It is not
difficult to double check that this vector satisfies equation
(\ref{null cone}) up to the first order in $\alpha$. To obtain the
expression for $k_i$ we use the following relation,
\begin{equation}
{\tilde k}^{i}=g^{i\mu}{\tilde
k}_{\mu}=-g^{i}k_{0}-\gamma^{ij}{\tilde k}_{j}
\end{equation}
which is a direct consequence of the relations given in (\ref{relation}).
Therefore
\begin{eqnarray}
{\tilde k}_{i}&=&-\gamma_{ij}\left({\tilde k}^{j}+k_0g^{j}\right)\nonumber\\
&=&-k_{0}\left[g_{i}+\gamma_{ij}\frac{1}{\sqrt{h}}
\frac{dx^j}{dl}\left(1-\frac{A_{00}}{2}D^2\right)-\frac{1}{2\sqrt{h}}\gamma_{ij}\left(
2A_{0k} \gamma^{kj}D +
A_{kl}\gamma^{lj}\frac{dx^k}{dl}\right)\right]\nonumber\\
&=&-k_0\left[g_{i}+\gamma_{ij}\frac{1}{\sqrt{h}}
\frac{dx^j}{dl}\left(1-\frac{A_{00}}{2}D^2\right)-\frac{1}{2\sqrt{h}}\left(
2A_{0i}D + A_{ki}\frac{dx^k}{dl}\right)\right] .
\end{eqnarray}
Multiplying by $dx^{i}$ and dropping the constant factor $k_{0}$
we end up with the following modified version of the Fermat
principle,
\begin{equation}
\delta\int{\frac{dl}{\sqrt{h}}(1-\frac
{A_{00}}{2}D^2)+(g_{i}-\frac{D}{\sqrt{h}}A_{0i})dx^i-\frac{1}{2\sqrt{h}}A_{ij}dx^i
\frac{dx^j}{dl}}=0 .\label{Fermat}
\end{equation}
In a static background (for which $D=\frac{1}{\sqrt{h}}$) we have
\begin{equation}
\delta\int{(1-\frac{A_{00}}{2h})\frac{dl}{\sqrt{h}}-\frac{A_{i0}dx^{i}}{h}-\frac{A_{ij}dx^{i}dx^j}{2\sqrt{h}dl}}=0
.\label{static}
\end{equation}
Comparison of (\ref{Fermat}) with Fermat's principle in its classical form, i.e  $\delta\int{n dl}=0$ \cite{9,10},
indicates that the trajectory of a light ray in a gravitational field is determined in the same way as in an
inhomogeneous refractive medium with $n$ as its {\it effective local index of refraction} given by
\begin{equation}
n=\frac{1}{\sqrt{h}}(1-\frac
{A_{00}}{2}D^2)+(g_{i}-\frac{D}{\sqrt{h}}A_{0i})e^{i}-\frac{1}{2\sqrt{h}}A_{ij}e^{i}e^{j}
,\label{refractive index}
\end{equation}
in which $e^{i}=\frac{dx^{i}}{dl}$ is the tangent vector to the
ray. The equations (\ref{Fermat}), (\ref{static}) and
(\ref{refractive index}) reduce to their corresponding classical
counterparts at zero $\alpha$ limit. An alternative representation
is given by introducing the four vector
$E^{\mu}=\frac{dx^\mu}{dl}$, which is null with respect to the
background metric and its components are $\left(D,e^{i}\right)$.
The covariant components of the tangent vector $e^i$ form a
three-dimensional vector in the space with metric $\gamma_{ij}$
and correspondingly the square of this vector is equal to one.
\begin{equation}
e_{i}=\gamma_{ij}e^{j}\ \
e^2=\gamma_{ij}e^{i}e^{j}=1
\end{equation}
With such a definition, the effective local index of refraction is
expressed as
\begin{equation}
n=E^{0}-\frac{1}{2\sqrt{h}}A_{\mu\nu}E^{\mu}E^{\nu}.
\label{refractive index1}
\end{equation}
This is obviously a result more general than the standard gravitational result
discussed in \cite{10} where only the effect of the gravitational field (in the context of $1+3$ approach), as a dispersive medium, on photon propagation is considered. A comparison between the classical
wave vector $k^{\mu}=\left(k^{0},\frac{k_0}{\sqrt{h}}e^{i}\right)$
and $E^{\mu}$ shows that $k^{\mu}=\frac{k_0}{\sqrt{h}}E^{\mu}$
i.e, these two vectors are parallel . Therefore we have
\begin{equation}\label{refractive index correction}
\delta n=-\frac{1}{2\sqrt{h}}A_{\mu \nu} E^{\mu}
E^{\nu}=-\frac{\sqrt{h}}{2k_{0}^2}A_{\mu \nu} k^{\mu} k^{\nu
}=-\frac{\sqrt{h}}{2k_{0}^2}(\delta k^2),
\end{equation}
where $\delta n$ and $\delta k^2$ are refractive index and null
cone modifications respectively. The latter has been studied for
different directions and polarizations
in many curved backgrounds \cite{3,4}. So evaluation of $\delta n$ reduces to the multiplication of $\delta k^2$
by the local factor $-\frac{\sqrt{h}}{2k_0^2}$.


\section{Polarization sum rules and horizon theorem}
In the previous section  it is shown that $\delta n$ is
proportional to   $\delta k^2$, using (\ref{refractive index
correction}) we can state {\em two theorems parallel to
polarization sum rules and horizon theorem in }\cite{8}.
\subsection{Polarization sum rules}
For Ricci flat space-times, the sum over the two physical
polarization of the refractive index is zero i.e,
$\sum\limits_{pol}\delta n= 0$, but for non Ricci flat space-times
satisfying the Einstein field equations,
\begin{equation}
\sum\limits_{pol}\delta n=\frac
{16\pi}{m^2}\frac{\sqrt{h}}{k_0^2}(b+2c){ T_{\mu
\nu}}k^{\mu}k^{\nu}.\label{sum n}
\end{equation}
Substituting for coefficients a and b ,we arrive at
\begin{equation}
\sum\limits_{pol}\delta n=\frac
{22}{45}\frac{\alpha}{m^2}\frac{\sqrt{h}}{k_0^2}{ T_{\mu
\nu}}k^{\mu}k^{\nu} ,\label{sum n1}
\end{equation}
where $T_{\mu\nu}$ is the energy momentum tensor. Since in the
weak field perturbative expansion, we can take $k^{\mu}$ to be
null on the right hand side of equation (\ref{sum n1}), the null
energy condition implies $T_{\mu\nu}k^{\mu}k^{\nu}\geq0$. So
assuming the null energy condition, we always have
$\sum\limits_{pol}\delta n \geq0$. This relation is consistent
with the polarization summed velocity shift,
$\sum\limits_{pol}\delta v\leq0$ \cite{8}.
\subsection{Horizon theorem}
In \cite{8}, it is shown that for spacetimes with an event horizon, photons with
momentum directed normal to the horizon have velocity equal to c,
i.e. the light cone remains $k^2=0$ independent of photon
polarization. Using (\ref{refractive index correction}) yields
that such (QGO) photons also have {\em local refractive index equal to $E^0$,
independent of their polarization}.\\
\section{Electromagnetic analogy}
With a similar discussion, one can find the modified
refractive index corresponding to the modified light
cone condition in an arbitrary anisotropic but homogenous electromagnetic field in flat space-time.
The description of such a photon based on the one-loop action can be given by an expansion in powers of $\alpha \frac{F^{2}}{m^4}$ which is a dimensionless parameter defined as the ratio of the background scale to a typical mass in the theory. Starting from the Euler-Heisenberg \cite{12} effective action
\begin{equation}
\Gamma=\int dx \left[-\frac{1}{4}F_{\mu\nu}F^{\mu\nu}+\frac{1}{m^4}\left(z\left(F_{\mu\nu}F^{\mu\nu}\right)^2+yF_
{\mu\nu}F_{\lambda\sigma}F^{\mu\lambda}F^{\nu\sigma}\right)\right]
\end{equation}
with $z=-\frac{1}{36}\alpha^2$ and $y=\frac{7}{90}\alpha^2$, we derive the equation of motion\\
Analogous to the equation (\ref{pol}) in the gravitational case,
using similar arguments the geometric optics
approximation gives
\begin{equation}
k_{\mu}f^{\mu\nu}-\frac{16z}{m^4}F^{\mu\nu}F_{\lambda\sigma}k_{\mu}f^{\lambda\sigma}-\frac{8y}{m^4}\left(F^{\mu\lambda}
F_{\lambda\sigma}k_{\mu}f^{\nu\sigma}+F^{\mu\lambda}F^{\nu\sigma}k_{\mu}f_{\lambda\sigma}\right)=0,
\end{equation}
where $F_{\mu\nu}$ is the background electromagnetic field strength. Rewriting the above equation as an equation for the
 physical polarization vector, $a^{\mu}a_{\mu}=-1$, we have
\begin{equation}
k^2+\frac{8}{m^4}\left(4z+y\right)F_{\mu\nu}F_{\lambda\sigma}k^{\mu}k^{\lambda}a^{\nu}a^{\sigma}-\frac{8y}{m^4}
F_{\mu}^{\ \nu}F_{\lambda\nu}k^{\mu}k^{\lambda}=0
\end{equation}
For a polarization state orthogonal to the plane spanned by the photon momentum and magnetic field direction,
in which $F_{\mu\nu}a^{\mu}=0$, the light cone condition is
\begin{equation}\label{ref1}
k^2=\frac{8y}{m^4}F_{\mu}^{\ \lambda}F_{\nu\lambda}F_{\mu\nu}k^{\mu}k^{\nu}=-\frac{8y}{m^4}T_{\mu\nu}k^{\mu}k^{\nu}
\end{equation}
Here in the second equality with the classical energy-momentum
tensor expression, $T_{\mu\nu}=-F_{\mu}^{\
\lambda}F_{\nu\lambda}+\frac{1}{4}\eta_{\mu\nu}F_{\lambda\sigma}F^{\lambda\sigma}$,
the higher order corrections are neglected. For the second
state coplanar with the above plane, it is given by
\begin{equation}\label{ref2}
k^2=-\frac{16}{m^4}\left(2z+y\right)T_{\mu\nu}k^{\mu}k^{\nu}
\end{equation}
where in both cases (\ref{ref1}) and (\ref{ref2}), the null cone dependence on the energy momentum tensor is similar to that of
the gravitational case and the polarization dependence appears
through the coefficient of $T_{\mu\nu}k^{\mu}k^{\nu}$ (or what
we have called $A_{\mu\nu}$ in equation (\ref{null cone0})). Setting $h$
and $g_\alpha$ equal to one and zero
respectively in defination of $E^{\mu}$ components and using equation (\ref{refractive index1}),
the local refractive index for each polarization is given by
\begin{eqnarray}
n_1&=&1+\frac{1}{k_{0}^2}\frac{4y}{m^4}T_{\mu\nu}k^{\mu}k^{\nu}\nonumber\\
n_2&=&1+\frac{1}{k_{0}^2}\frac{8}{m^4}\left(2z+y\right)T_{\mu\nu}k^{\mu}k^{\nu}
\end{eqnarray}
Again notice that if $T_{\mu\nu}k^{\mu}k^{\nu}\geq0$, both
polarizations have refractive indices larger than one. It is also notable that the above expressions agree with the results of \cite{Tsai} in the weak field and low frequency limit. Now taking
the polarization sum and substituting for the explicit values of 
coefficients $y$ and $z$, we find
\begin{equation}
\sum\limits_{pol}\delta n=\frac{11}{45}\frac{\alpha^2}{m^4 k_0^2}T_{\mu\nu}k^{\mu}k^{\nu}
\end{equation}
This should be compared with eq. (\ref{sum n1}) for the gravitational case. Notice that for electromagnetism, the sum over the one-loop correction is proportional to $\alpha^2$, essentially because the lowest order terms in the Euler-Heisenberg action are of 4th order in the background fields, i.e. $O\left(F^4\right)$, compared to the 3rd order terms (i.e. $O\left(RFF\right)$) in the Drummond-Hathrell action \cite{2}.

\section{Spatial Light Paths}
The stationarity of the photon arrival time under the first
order variation of $\gamma$ spatially translates into the stationarity of
its arbitrarily parametrized spatial arc length
\begin{equation}\label{ind0}
\delta \int n(x^i, e^i)\frac{dl}{d\lambda}d\lambda= \delta \int n(x^i, {\hat{e}}^i)\sqrt {{\hat{e}}^i{\hat{e}}_i} d\lambda=0,
\end{equation}
in which $\lambda$ is a general parameter and  ${\hat{e}}^i = e^i\frac{dl}{d\lambda}$.
For an stationary spacetime the above variation reduces to the spatial light path equation 
\begin{equation}\label{ind1}
\frac{\partial}{\partial x^i}{\hat{n}}=\frac{d}{d\lambda}\left(\frac{\partial {\hat{n}}}{\partial {\hat{e}}^i}\right)
\end{equation}
in which ${\hat{n}}= n(x^i, {\hat{e}}^i)\sqrt{{\hat{e}}^i{\hat{e}}_i}$. Now choosing $\lambda$ for the extremal path to be proportional to the spatial arc length $l$, i.e $\lambda = \alpha l$, we end up with the following equation for the 
light paths
\begin{equation}\label{ind}
\frac{\partial}{\partial x^i}(n \sqrt {{e}^i{e}_i}) =\frac{d}{dl}\left(\frac{\partial (n \sqrt {{e}^i{e}_i})}{\partial {e}^i}\right)
\end{equation}
Now for the classical part of $n$, i.e in the absence of the quantum corrections, we have
\begin{eqnarray}
\frac{d}{dx^i}(E^0 \sqrt {{e}^i{e}_i})&=&\partial_i \frac{1}{\sqrt{h}}+e^j\partial_i g_j\nonumber\\
\frac{d}{dl}\left(\frac{\partial (E^0 \sqrt {{e}^i{e}_i})}{\partial
e^i}\right)&=&\frac{d}{dl}\left[\frac{1}{\sqrt{h}}e_i+ g_i+ \left(g_j e^j\right)e_i
\right]\nonumber\\
&=&e_i\left(\stackrel{\rightarrow}{e}.\stackrel{\rightarrow}{\nabla}\right)\frac{1}{\sqrt{h}}+\left(\frac{1}{\sqrt{h}}+g_j e^j\right)
\frac{d}{dl}e_i+\left(\stackrel
{\rightarrow}{e}.\stackrel{\rightarrow}{\nabla}\right)g_i\nonumber\\
\end{eqnarray}
After some manipulation we get 
\begin{eqnarray}
E^0\frac{d}{dl}e_i&=& \left(\nabla_\bot\right)_i \frac{1}{\sqrt{h}}+e^j
\left(\partial_i g_j-\partial_j g_i\right)\nonumber\\
&=&\left(\nabla_\bot\right)_i\frac{1}{\sqrt{h}}+\left[\stackrel{\rightarrow}{e}\times
\left(\stackrel{\rightarrow}{\nabla}\times\stackrel{\rightarrow}{g}
\right)\right]_i \label{classic spatial}
\end{eqnarray}
where
$\nabla_\bot\equiv\nabla-\stackrel{\rightarrow}{e}\left(\stackrel{\rightarrow}{e}.\nabla\right)$
denotes the projection of $\nabla$ onto the plane normal to the direction of the ray given by the unit tangent vector $\stackrel{\rightarrow}{e}$. This equation describes how a classical light ray
in a curved stationary background is deviated from a straight line in Minkowski space. Deviation is devided into two intrinsically different parts. The first term corresponds to an attraction towards the deflection center, the so called {\it gravitoelectric} force, whereas the second term is due to the {\it gravitomagnetic} force originated from the gravitomagnetic potential $\stackrel{\rightarrow}{g}$ \cite{lynd}.  \\
In the presence of the quantum interactions, the equation  for the spatial light paths, eq.(\ref{ind}), is given by
\begin{equation}
\partial_i E^0-\partial_i\left(\frac{\hat{A}_{\mu\nu}}{2\sqrt h}E^\mu E^\nu\right)=\frac{d}{dl}\left[\frac{\partial}{\partial e^i}E^0-\frac{\partial}{\partial e^i}\left(\frac{\hat{A}_{\mu\nu}}{2\sqrt h}E^\mu E^\nu\right)\right]
,\label{spatial path correction}\end{equation}
where $\hat{A}_{\mu\nu}={A}_{\mu\nu}\sqrt{{{e}}^i{{e}}_i}$ and the correction term in right hand side is
\begin{eqnarray}
&&\frac{d}{dl}\left[\frac{{A}_{\mu\nu}}{2\sqrt h}E^\mu E^\nu e_i+\frac{\hat{A}_{\mu\nu}}{\sqrt h}\left(\frac{\partial}{\partial e_i} E^{\left\{\mu\right.}\right)E^{\left.\nu
\right\}}\right]=e_i\left(\stackrel{\rightarrow}{e}.\stackrel{\rightarrow}{\nabla}\right)\left(\frac{{A}_{\mu\nu}}{2\sqrt h}E^\mu E^\nu\right)+\frac{{A}_{\mu\nu}}{2\sqrt h}E^\mu E^\nu\frac{d}{dl}e_i\nonumber\\
&&+\left(\stackrel{\rightarrow}{e}.\stackrel{\rightarrow}{\nabla}\right)\frac{{A}_{\mu\nu}}{\sqrt h}\left(\frac{\partial}{\partial e^i }E^{\left\{\mu\right.}\right)E^{\left.\nu\right\}}+\frac{A_{\mu\nu}}{\sqrt{h}}\left[\frac{d}{dl}\left(
\frac{\partial}{\partial e^i} E^{\{\mu}
\right)E^{\nu\}}+\left(\frac{\partial}{\partial e ^i} E^{
\{\mu}\right) \frac{d}{dl} E^{\nu\}} \right]
.\label{secod term}\end{eqnarray}
Gathering all together, we have
\begin{eqnarray}
&&\frac{1}{n}\frac{d}{dl}e_{i}=\left(\stackrel{\rightarrow}{\nabla}_{\bot}\right)_{i}\left(\frac{1}{\sqrt h}-\frac{A_{\mu\nu}}{2\sqrt h}E^{\mu} E^{\nu}\right)+\left[\stackrel{\rightarrow}{e}\times\left(\stackrel{\rightarrow}{\nabla}\times \stackrel{\rightarrow}{g}\right)\right]_{i}\nonumber\\
&&+\left(\stackrel{\rightarrow}{e}.\stackrel{\rightarrow}{\nabla}\right)\frac{{A}_{\mu\nu}}{\sqrt h}\left(\frac{\partial}{\partial e^i }E^{\left\{\mu\right.}\right)E^{\left.\nu\right\}}+\frac{A_{\mu\nu}}{\sqrt{h}}\left[\frac{d}{dl}\left(
\frac{\partial}{\partial e^i} E^{\{\mu}
\right)E^{\nu\}}+\left(\frac{\partial}{\partial e ^i} E^{
\{\mu}\right) \frac{d}{dl} E^{\nu\}} \right]
, \label{spatial path}\end{eqnarray}
in which $A^{\left\{\mu\right.}B^{\left.\nu\right\}}=\frac{1}{2}\left(A^{\mu}B^{\nu}+A^{\nu}B^{\mu}\right)$ and the derivatives are given by the following relations in the same coordinate frame
\begin {eqnarray}
\partial_i E^\mu&=&\left(\partial_i\left(\frac{1}{\sqrt{h}}\right)+e^j\partial_i g_j, \stackrel{\rightarrow}{0}\right)
\nonumber\\
\frac{\partial}{\partial e^i}E^\mu&=&\left(g_i,\delta^{j}_i\right)\nonumber\\
\frac{d}{dl}\frac{\partial}{\partial e^i}E^\mu&=&\left(\stackrel{\rightarrow}{e}.\stackrel{\rightarrow}{\nabla}g_i,\stackrel{\rightarrow}
{0}\right)\nonumber\\
\frac{d}{dl}E^\mu&=&\left(\stackrel{\rightarrow}{e}.\stackrel{\rightarrow}{\nabla}\left(\frac{1}{\sqrt
{h}}\right)+\stackrel{\rightarrow}{e}.\stackrel
{\rightarrow}{\nabla}\left(g_j\right)e^j+g_{j}\frac{d}{dl}e^j,\frac{d}{dl}\stackrel{\rightarrow}{e}\right).
\end{eqnarray}
The equation (\ref{spatial path}) is a complicated combination of
curvature and metric components which is formulated through the
stationarity of the arrival time under the variations of the null
curves connecting the source to the timelike observer in the presence of quantum corrections. From the above calculations, compared to the classical path, the time delays are expected to be of order $\alpha$ . Therefore the natural
question one should address is whether these corrections pile up
as the travel distance increases. If the answer is yes, then a
deviation much larger than the scale of the theory could be
achieved. A more detailed discussion of this topic will be given
elsewhere \cite{11}.
\section{Discussion}
It is natural to expect that the inclusion of QED vaccum polarization effect on photons propagating in a background gravitational field should modify their trajectories in that field. This is actually part of the ongoing studies under the general title of quantum gravitational optics (QGO).
To look at the roots of this trajectory modification we have studied Fermat's
principle
in QGO in the context of threading ($1+3)$ formulation of spacetime decomposition which is best adapted to the employed version of the principle. Taking the above QGO effects into account, it is shown there is an effective
Fermat's principle for the propagation of a ray in a general stationary background through the introduction of an explicit expression for the {\it effective
index of refraction}. Obviously not only QGO photons do not propagate along the shortest line
in space but also the effective refractive index is itself a complicated combination of curvature and metric components as well as the quantum corrections considered in the context of QGO.
\\The modification of the refractive index is shown to be proportional to
the null cone modification in the presence of QGO effects i.e $\delta k^2$. This fact is used to show that the polarization summed shift in the refractive indices is greater than or equal to zero. It is also shown that for photons with momentum directed normal to the horizon the refractive index being equal to the classical case is polaraization independent.\\
Polarization dependent refractive indices for the electromagnetic analogue of the same problem are calculated explicitly and compared with their gravitational counterparts in the weak field and low frequency limit. Finally, through the effective refractive index, the equation of spatial light paths in the presence of the above quantum corrections is found.
\section*{Acknowledgment}
N. A and M. N-Z thank University of Tehran for supporting this project under the grants provided by the research council.
M. N-Z also thanks the center of excellence in the structure of matter for the partial support.
\bibliographystyle{amsplain}

\end{document}